\documentclass{iopjournal}
\usepackage[
backend=biber,
style=numeric-comp,
sorting=none
]{biblatex}
\usepackage{xcolor}
\newcommand{\mvh}[1]{\textcolor{blue}{#1}}
\newcommand{\mht}[1]{\textcolor{olive}{#1}}
\usepackage{amsmath}

\begin{document}

\articletype{Paper} 

\title{Interacting Hysterons with Asymptotically Small or Large Spans
}

\author{Margot Teunisse$^{1,2}$
and Martin van Hecke$^{1,2}$
}

\affil{$^1$AMOLF, Amsterdam, The Netherlands}

\affil{$^2$LION, Leiden University, Leiden, The Netherlands}

\email{margotteunisse@outlook.com}


\begin{abstract} 
Models of interacting hysteretic elements, called hysterons, capture the sequential response and complex memory effects in a wide range of complex systems and can guide the design of intelligent metamaterials. However, even simple models with few hysterons feature a bewildering number and variety of behaviors. Here we study the hysteron model in two physically relevant limits, where {the} response {of a hysteron system} is easier to understand. First, when the hysteron span - the gap between its two hysteretic transitions - dominates all other scales, the range of pathways encoded in transition graphs (t-graphs) becomes limited because many avalanches {are} absent. Second, when the hysteron span becomes vanishingly small, hysterons behave as interacting binary spins, {which require avalanches in order to} exhibit nontrivial pathways. Finally we show that hysterons can be 
mimicked by pairs of strongly interacting spins, {such} that
collections of $n$ interacting hysterons can be mapped to $2n$ interacting spins, albeit {via} highly specific interactions. {Altogether,} our work provides a deeper understanding of the role of the hysteron parameters on their collective behavior, and points to connections and differences between spin- and hysteron-based models of complex matter.
\end{abstract}

\section{Introduction}

Multistable, glassy materials, ranging from
crumpled sheets and 
frustrated magnets to metamaterials, 
irreversibly transition between metastable states when externally driven \cite{keim2019memory}. 
In the limit of low temperature and for slow homogeneous driving, such as mechanical compression, shear, or a magnetic field, these pathways can be described by transition graphs (t-graphs) \cite{baconnier2025dynamic,bense2021,colintoggleron2025,ding2022sequential,ferrari2022preisach,jerryseriallycoupled,jin2025dynamic,juleslechenault2021,lindeman2021,lindeman2023competition,lindeman2025generalizing,melancon2022,muhaxheri2024bifurcations,muhaxheri2025catastrophic,mungan2019structure,paulsen2024mechanical,regev2019,shohat2022memory,shohat2025geometric,keimpaulsen2019,terzimungan2020,vhecke2021}. The nodes of these graphs represent the metastable states, and their edges represent the irreversible transitions that occur when the driving exceeds certain critical values. 
Such t-graphs can be extracted from experimental data \cite{bense2021,jin2025dynamic,colintoggleron2025,jerryseriallycoupled,ding2022sequential} or numerical simulations of more realistic models \cite{shohat2025geometric,jerryseriallycoupled,regev2019}, and provide a full description of the materials pathways under any complex driving protocol. Hence, the structure of these graphs can be used to gain  insight into memory effects, where the state encodes a memory of past driving.

Physically,  many metastable systems can be thought of as collections of interacting bistable elements. These can take the form of directly observable units, such as slender beams or snapping creases \cite{kwakernaak2023,shohat2022memory}, or can be emergent, such as the two groups of local particle configurations that separate T1 reorganizations of amorphous media \cite{keimpaulsen2021}.
{The} decomposition {of multistable systems} into bistable units
has inspired work to model the response of such systems with hysterons. These are spin-like, minimalistic two-state elements , described by a binary phase {$s_i \in \{0, 1\}$,}
{which} exhibit hysteresis. An isolated hysteron $i$ flips from $s_i= 0$ to $s_i=1$ when the global driving{, denoted $U$,} exceeds the up switching threshold
$u_i^+${; similarly, it flips from $s_i = 1$ to $s_i=0$}  when $U$ falls below the down switching threshold $u_i^-$, with $u_i^+> u_i^-$ \cite{vhecke2021,keimpaulsen2021,lindeman2021,regev2019}.
Ensembles of non-interacting hysterons form the so-called Preisach model, which has been studied intensely as it {provides an explanation for return point memory (RPM)}\cite{terzimungan2020,mungan2019structure,deutsch,mungan2019structure}. For many physical systems, {however}, the bistable elements are coupled, for example through the elastic background in which they are embedded. {This suggests that models describing these systems should incorporate interactions between hysterons}\cite{bense2021,shohat2022memory,shohat2025geometric,liu2024,lindeman2021,keimpaulsen2021}. 
{The hysteron} interactions can be expressed via 
state-dependent switching thresholds $U_i^\pm(S)$,
where $S:=(s_1,s_2,\dots)$. Here we consider
pairwise interactions: 
\begin{equation}
    U_i^\pm(S) = u_i^\pm - \sum_{j}c_{ij}s_j~,
\end{equation}
where $u_i^\pm$ are the bare switching thresholds, 
the coupling coefficients $c_{ij}$ express the effect of the phase of hysteron $j$ on the thresholds of hysteron $i$, and we gauge out the self-coupling $(c_{ii}=0)$ \cite{vhecke2021}.

For three non-interacting hysterons, only six t-graphs are possible; more generally, $n$ non-interacting hysterons can produce $n!$ distinct t-graphs. \cite{terzimungan2020,mungan2019structure,regev2019,ferrari2022preisach,preisach1935,vhecke2021,juleslechenault2021}. 
Hysteron interactions give rise to a much wider range of t-graphs. Numerical sampling has revealed that the number of t-graphs for three pairwise interacting hysterons (far) exceeds $15\times10^3$ \cite{vhecke2021}. These t-graphs {moreover} encode wide range of behaviors and memory effects, including transients and multiperiodic responses under cyclic driving \cite{paulsen2024mechanical,keimpaulsen2019,szulc2022cooperative}, latching \cite{lindeman2025generalizing}, and even computing \cite{kwakernaak2023,liu2024}, and  understanding how such effects arise requires  examining the structure and multiplicity of t-graphs in the hysteron model.

Here we consider the hysteron model in two specific, physically motivated limits, where {the} complexity {of the model} can be partly tamed. We define the {\em span} $\sigma_i$ of a hysteron as the difference between the bare up and down thresholds: $\sigma_i \equiv u_i^+ - u_i^-$, and consider the limits of small and large span. We first consider
the limit of large spans, {where} $U_i^+(S)>U_j^-(S')$ for all $i,j$ and states $S$ and $S'$ (section.~\ref{ss:largespan}).
{We then} consider the limit of small span, where the hysterons act as binary spins (section.~\ref{ss:spin}). The up and down transitions between two collective states that differ only in the phase of a single spin then become equal. We show that this severely limits the range of t-graphs, and that all non-trivial graphs need to feature avalanches.
Finally, we show that pairs of interacting spins can 
mimic hysterons (section~\ref{ss:spinpair}), and provide a construction where any set of 
$n$ interacting hysterons can be mapped to a set of $2n$ interacting hysterons (section~\ref{ss:rep}).
{Altogether}, our work provides examples of limits where the hysteron model is easier to study, as well as a firm link between models of interacting hysteron and interacting spin.

\section{Pairwise interacting hysterons}
In this section we briefly review the interacting hysteron model and the links between t-graphs and switching tresholds.
We model bistable elements in a physical system by 
collections of strictly binary hysterons, and model their
interactions by making the switching thresholds of each hysteron dependent on the phase of the other hysterons. Defining the binary state $S:=(s_1,s_2,\dots)$, we thus introduce the state-dependent switching thresholds $U_i^\pm(S)$ \cite{vhecke2021, teunisse, lindeman2021, keimpaulsen2021}. In principle, (a subset of) these thresholds can be measured experimentally, obtained from simulations of an underlying physical model, or in some cases, calculated explicitly \cite{bense2021,liu2024,shohat2025geometric}. Here we will not consider such a detailed connection, but instead simply pose that there are pairwise interactions with the strength of these interactions {being} equal for the up and down thresholds. {The switching thresholds can then be expressed as}
\begin{equation}
    U_i^\pm(S) = u_i^\pm - \sum_{j \neq i}c_{ij}s_j~.
\end{equation}

We recently {described} the connections between the switching thresholds and the t-graphs in detail~\cite{teunisse}, {building on earlier work~\cite{mungan2019structure, keimpaulsen2021,vhecke2021}.} {Below, we summarize the main results which are relevant to this paper.}
{These results consider} the mapping from switching thresholds $U_i^\pm(S)$ to the t-graph, and the inverse problem {of finding constraints on the switching thresholds for a given t-graph topology.}

\paragraph{Mapping from {switching thresholds} to t-graph.---}
The switching thresholds determine the stability range of each state $S$: the upper state threshold
$U^+(S):= \mbox{min}_{i_0} U_i^+(S)$, where $i_0$ are the indices of the hysterons in phase '0', and 
the lower state threshold
$U^-(S):= \mbox{max}_{i_1} U_i^-(S)$, where $i_1$ are the indices of the hysterons in phase '1'. Similarly, these thresholds determine which 'critical' hysteron will flip when a state becomes unstable due to an increase or decrease of the driving $U$. The collection of states and their critical hysterons forms the {\em scaffold}, which can be seen as {the basis upon which a t-graph is built \cite{teunisse}.

To go from a scaffold to a t-graph, one must consider the nature of the transitions that occur
once a state $S$ is destabilized. When state $S$ becomes unstable as the driving $U$ is swept up (down), the critical up (down) hysteron flips, and the system reaches an intermediate state $S'$, as determined by the scaffold. If the state $S'$ is stable at the driving $U$ where the transition was initiated, a simple transition occurs between two states that are separated by one hysteron flip. If this state is unstable, avalanches occur; for these avalanches we make a distinction between cases where either a single hysteron or multiple hysterons is (are) unstable}. In the first case, where only one hysteron is unstable in state $S'$ at driving $U$, flipping that hysteron produces a state $S''$ - which again can be stable or not, etc. In the second case, where more than one hysteron is unstable in state $S'$ at driving $U$, the system is in a race condition -- i.e., the order of hysteron flips is ambiguous \cite{vhecke2021,teunisse,baconnier2025dynamic}. One can then deem the system ill-defined \cite{vhecke2021}, or introduce a phenomenological rule such as 'flip the most unstable hysteron' to proceed \cite{baconnier2025dynamic,keimpaulsen2021}.

Given a specific rule to resolve race conditions, one can determine the collection of transitions - simple and avalanches - that occur when $U$ is swept, and collect the resulting states and transitions in a transition graph (t-graph). 
For large systems with random interactions, self-loops can occur; in this paper we consider such systems ill-defined \cite{baconnier2025dynamic}. 
We note that each transition follows the scaffold, and that each valid avalanche can be composed of a number of elementary hysteron flips consistent with the scaffold; hence, all t-graphs can be seen as scaffolds dressed by avalanches \cite{teunisse}.

\paragraph{Mapping from t-graph to {switching thresholds}.---}
{We and others have previously considered the inverse problem of constructing constraints on the switching thresholds such that a given t-graph is realized~\cite{vhecke2021, teunisse,keimpaulsen2021,liu2024,shohat2025geometric,muhaxheri2024bifurcations}.} These constraints take the form of inequalities between the switching thresholds, which determine the range of stability of states, the scaffold, and the presence of avalanches \cite{vhecke2021,teunisse}. For a given t-graph topology, the full set of inequalities {is} referred to as the design inequalities; these  can be solved using standard techniques{. Moreover,} for a given {t-graph} topology one can {easily determine} if there is a set of design parameters that can produce this graph --- for details see \cite{teunisse}.

{We note that} all design inequalities are of the form $U_i^\pm(S_A) > U_j^\pm(S_B)$. {For the purposes of this paper,} it is convenient to
{classify}
{the design} inequalities into two {types}. The first {type compares} up with up, or down with down {switching thresholds}. {Provided that interactions are pairwise, we can write} 
\begin{eqnarray}
    U_i^+(S_A)-U_j^+(S_B) &=& (u_i^+ - u_j^+) - f(S_A, S_B, c_{ij}) > 0~,\label{eq:upup}\\
U_i^-(S_A)-U_j^-(S_B) &=& (u_i^- - u_j^-) - f(S_A, S_B, c_{ij}) > 0~\label{eq:downdown},
\end{eqnarray}
where $f(S_A, S_B, c_{ij})$ is a linear combination of the coupling coefficients:
\begin{equation}
f(S_A, S_B, c_{ij}) = \sum_{k \neq i}c_{ik}s_{k, A} - \sum_{k \neq j}c_{jk}s_{k, B}~,
\end{equation}
and where $s_{k, A}, s_{k, B}$ are the phases of hysteron $k$ in state $S_A$ and $S_B$, respectively.
The second group contains the inequalities that compare up and down thresholds:
\begin{equation}
    U_i^+(S_A)-U_j^-(S_B) = (u_i^+ - u_j^-) - f(S_A, S_B, c_{ij}) > 0~.\label{eq:updown}
\end{equation}
{We refer to the inequalities expressed by Eqs.~\ref{eq:upup}-\ref{eq:downdown} and Eq.~\ref{eq:updown} as type-I and type-II inequalities, respectively.}

For each transition
{$S^0 \rightarrow S^l$, which evolves through intermediate states
$S^1,\dots, S^{l-1}$,} we distinguish three groups of inequalities. 
The start of {the} transition, $S^0 \rightarrow S^1$, {is determined by a group of inequalities which we refer to as the initial inequalities}. {The initial inequalities} depend on 
the critical hysteron of state $S^0$ as encoded by the scaffold{; moreover, they require that the state $S^0$ is initially stable}. {The inequalities that determine the scaffold are of type I, while the inequalities which determine the initial stability of state $S^0$ are of type II.}

The stability of the intermediate states, and {of} their critical hysterons, are determined by a mix of {type-I and type-II inequalities; we refer to these as} the intermediate inequalities.
Finally, the stability of the final states again depends ib
a mix of {type-I and type-II inequalities}, which we refer to as the final inequalities. 

{In summary, the scaffold depends on type-I inequalities, the stability of the initial state depends on type-II inequalities, and avalanches depend on a mix of type-I and type-II inequalities. We make use of this classification when we consider the impact of the span on a t-graph.}

\section{Large Span}\label{ss:largespan}

{We analyze t-graphs of pairwise interacting hysterons with a fixed interaction matrix \(c_{ij}\), and consider the effect of increasing the span $\sigma_i$. Specifically, we consider
additive changes of the span, where
we, e.g., map $u_i^- $ to $u_i^- - \Delta\sigma/2$ and
$u_i^+ $ to $u_i^+ + \Delta\sigma/2$ -- here $\Delta\sigma$ is a positive parameter that increases the spans of all hysterons equally. {By changing the spans in this way, we change the mean span $\langle \sigma_i \rangle$ to $\langle \sigma_i \rangle + \Delta\sigma$, while keeping the scatter $u_i^+ - u_j^+$ and $u_i^- - u_j^-$ of the bare switching thresholds fixed.}

{Crucially, we now note that only type-II inequalities (Eq.~\ref{eq:updown}) feature the mean span $\langle \sigma_i \rangle$; features which only depend on type-I inequalities (Eqs.~\ref{eq:upup}-\ref{eq:downdown}) are invariant to changes in the mean span. We make use of this insight to describe the effect of large mean span on the scaffold, state stability, and avalanches in a t-graph.}

\paragraph{State stability.---}
The range of stability of a given state $S$
depends on the difference between the lowest
$U_i^+(S)$ and highest $U_i^-(S)$. For a given choice of parameters, this range may become negative for some states; we refer to such states as being persistently unstable \cite{teunisse}. {Since these are type-II inequalities, as we have noted above, the range of stability of state $S$ is dependent on $\Delta\sigma$. Namely,} increasing the spans for given $c_{ij}$
eventually stabilizes all such states: if 
the mean span $\sigma$ is sufficiently large compared to $|u_i^+-u_j^+|$, $|u_i^- - u_j^-|$ and $|c_{ij}|$ for all $i, j$, then $U_i^+(S_A)>U_j^-(S_B)$ becomes true for all $i,j$ and choices of state, so that each state has a finite range of stability. 

\paragraph{Scaffold.---}
For the scaffold, the design inequalities are of the form $U_i^+(S) > U_j^+(S)$ and $U_i^-(S) > U_j^-(S)$; these inequalities specify that the critical up and down hysterons are those hysterons with the highest up and lowest down switching thresholds, respectively. {Since these are type-I inequalities, as we have noted above,} the scaffold is invariant to
the value of $\Delta\sigma$.

{\paragraph{Avalanches.---}
We now consider how an increase in the span impacts
avalanches. Each avalanche is labeled 'up' or 'down' depending on whether it is triggered by an up or down flip of a hysteron in response to an increase or decrease of $U$. 
We classify avalanches as monotonic when they 
consist of up or down flips only, and mixed otherwise.
} {The impact of an increase in the span on avalanches is nuanced: whereas the scaffold only depends on type-I inequalities and the stability range of the states $S$ only depends on type-II inequalities, avalanches depend on a mix of type-I and type-II inequalities. As we will show,} increasing the span does not affect monotonic avalanches, and truncates 
mixed avalanches so that only the monotonic initial part remains.

{Suppose that for $\Delta\sigma=0$, the t-graph of a set of interacting hysterons features
a monotonic avalanche. For definiteness we focus on a monotonic up avalanche $S^{0}\uparrow S^{1}\uparrow \dots S^{l}$, and note that the argument for down avalanches follows by symmetry. As the scaffold is not affected by $\Delta$, we only need to consider the intermediate and final inequalities. }

{The final inequalities are
	\begin{eqnarray}	    
		U^+(S^{0}) < U^+_{i}(S^{l}) ~~~ &\forall i\in I_0(S^{\lambda})~, \label{eq:final_span_invariance_1}\\
		U^+(S^{0}) > U_{i}^-(S^{l}) ~~~ &\forall i\in I_1(S^{\lambda})~, \label{eq:final_span_invariance_2}
	\end{eqnarray}
where  we note that 
the value of $U$ during the avalanche is given by the 
up switching threshold of state $S^{0}$.
Since both inequalities are true for 
$\Delta\sigma=0$, they remain true for positive $\Delta\sigma$; both terms in the first inequality increase by $\Delta\sigma/2$, whereas in the second inequality, the left term increases and the right term decreases for increasing $\Delta\sigma$. }

{The intermediate inequalities are
\begin{eqnarray}	    
U^+(S^{0}) &<& U^+_i(S^{\lambda}) ~ \forall i\in I_0(S^{\lambda})\backslash\{k\}~,\label{eq:interm_span_invariance_1}\\
U^+(S^{0}) &>& U_i^-(S^{\lambda}) ~ \forall i\in I_1(S^{\lambda})~, \label{eq:interm_span_invariance_2}\\
U^+(S^{0}) &>& U_k^+(S^{\lambda}) ~, \label{eq:interm_span_invariance_3}
\end{eqnarray}
where $k$ is the critical hysteron of state $S^\lambda$.
Following the same arguments as before, if these inequalities are satisfied for $\Delta\sigma=0$, they are also true for any positive $\Delta\sigma$. Hence, increasing the span does not affect monotonic avalanches\footnote{We point out that an increase in the span can resolve some race conditions. Suppose that an intermediate state for a monotonic up avalanche experiences race conditions. If these race conditions are such that one hysteron is unstable to up flips, and other hysterons are unstable to down flips, then an increase in the span can stabilize the latter group of hysterons.}}

{Now we consider mixed up avalanches, where after one or more up flips, there is a down flip. We label the intermediate state where this happens $S^\lambda$, and the relevant hysteron $k$.
This avalanche implies that for $\Delta\sigma=0$, the following inequality holds:
\begin{equation}
U^+(S^{0}) < U_k^-(S^{\lambda}) ~.
\end{equation}
When $\Delta$ is increased sufficiently, this inequality becomes false,
with  
state $S^\lambda$ becoming stable at $U^+(S^{0})$. Hence, the avalanche is truncated. For example, a mixed avalanche
of the form $S^0 \uparrow S^1 \uparrow S^2 \downarrow S^3 \dots $ is truncated to $S^0 \uparrow S^1 \uparrow S^2$ for large $\Delta$; similarly, a mixed avalanche of the form
$S^0 \uparrow S^1 \downarrow S^2 \dots $ is truncated to $S^0 \uparrow S^1$ and thus ceases to be an avalanche.}

{In summary, an increase in the span via the parameter $\Delta\sigma$ eventually stabilizes persistently unstable states, and truncates all mixed avalanches. For example, in systems that derive from serially coupled elements, we have shown that all 
avalanches are mixed and of length two \cite{liu2024} --- an increase in the span of these elements eventually suppresses all avalanches.
More generally, pairwise-interacting hysterons where all interactions are negative can only feature purely alternating avalanches (up-down-up-\dots); these will all be truncated to simple non-avalanche transitions for large $\Delta\sigma$. 
Thus, systems with purely negative coupling and large span can be employed in order to physically realize systems which feature scrambling without avalanches.}

\section{The zero-span spin limit}\label{ss:spin}

\begin{figure}
\centering
\includegraphics[width=.9\textwidth]{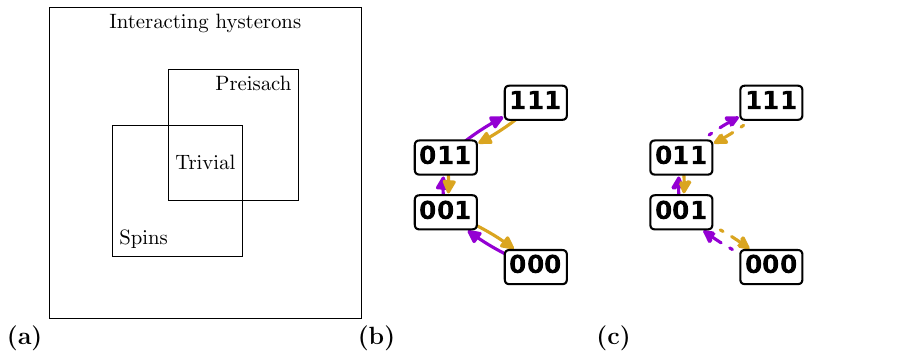}
\caption{{The spin limit of the hysteron model in the presence and absence of interactions. a) Schematic representation of spins within the parameter space of interacting hysterons: the space of interacting hysterons encompasses that of interacting spins, and the behaviour of independent spins is trivial. b) Non-interacting spins can only exhibit trivial t-graphs, where up and down transitions for each spin are paired. c) Scaffold for independent spins.}}
\label{fig:spin_parspace}
\end{figure}

{In this section we investigate the zero-span limit of hysteron systems -- we refer to such zero-span hysterons as 'spins' -- and in particular discuss the relations between interacting
hysterons and interacting spins. 
Such spin systems follow from the hysteron model by taking the upper and lower switching thresholds of 
hysteron $i$ in state $S$ to be identical: 
\begin{equation}
    U_i^+(S)=U_i^-(S)=:U_i(S)~.
\end{equation}
Note that these are zero-temperature, deterministic models - distinct from, e.g., spin glasses at finite temperature that are defined by a Hamiltonian. }

{Here we first investigate
the scaffolds and t-graphs of such systems, and show that 
scaffolds and avalanches are strongly intertwined for spins.
We then show that interacting spins can feature collective hysteretic behavior, with pairs of interacting spins being able to mimic hysterons. We show that all systems of $n$ interacting hysterons can be mimicked by systems of $2n$ interacting spins; to make this mapping, a careful consideration of race conditions is required. Hence, while hysteron models 
encompass models of interacting spins, 
in some sense, spin models 
also encompass models of interacting hysterons.}

\subsection{Scaffolds and avalanches for interacting spins---.}

{In the presence of interactions, interacting spin systems can show complex t-graph responses which, as for hysterons, can be categorized by separately considering the scaffold and avalanches \cite{teunisse}. We will show that any scaffold which can be realized by pairwise interacting hysterons can also be found for interacting spins. However, we will find that any non-trivial scaffold for interacting spins will always feature avalanches. As a result, any non-trivial t-graph without avalanches requires hysterons.}

{\paragraph{Scaffolds}
Here we discuss the scaffolds for interacting spins.
We first recall that for $n$ independent hysterons--i.e., the Preisach model--there are $n!$ possible t-graphs~\cite{terzimungan2020, ferrari2022preisach}. In contrast, for independent {\em spins}, the response to driving is trivial: the spins are simply in phase 0 or 1 if their switching threshold is above or below the current value of the driving $U$, respectively, so only a single 'trivial' t-graph is possible (Fig.~\ref{fig:spin_parspace}b). Accordingly, there is only a single possible trivial scaffold (Fig.~\ref{fig:spin_parspace}c). 
In the presence of interactions, however, all scaffolds that are realizable by pairwise interacting hysterons can also be realized by pairwise interacting trivial spins. }

{We first show how interacting spins can realize the Preisach scaffolds associated with non-interacting hysterons. We do so by establishing an explicit mapping between the parameters of the Preisach model and the interacting spin model, and showing that the design inequalities that govern their scaffolds are identical.  The equations that govern the scaffold compare the up and down switching thresholds of a given state. For all states $S$ where $s_i=0$ and $s_j=0$, we need to compare the up switching thresholds to determine the critical scaffold, i.e., 
the signs of 
the expressions $U_i^+(S) - U_j^+(S)$:
\begin{equation}
    U_i^+(S) - U_j^+(S) = (u_i^+ - u_j^-) - \sum_{k \neq i, j}(c_{ik} - c_{jk})s_k > 0~, \label{des1}
\end{equation}
where we made use of $s_i=s_j=0$ to write the last term. Similarly,
for all states $S$ where
$s_i=1$ and $s_j=1$, we need to compare the down switching thresholds to determine the critical scaffold, i.e., 
the signs of 
the expressions $U_i^-(S) - U_j^-(S)$:
\begin{equation}
U_i^-(S) - U_j^-(S) = (u_i^+ + c_{ji} - u_j^- - c_{ij}) - \sum_{k \neq i, j}(c_{ik} - c_{jk})s_{k} > 0~,\label{des2}
\end{equation}
where we use $s_i=s_j=1$ to 
separate the terms $c_{ij}$ and $c_{ji}$ from the sum. 
We note that the terms $\sum_{k \neq i, j}(c_{ik} - c_{jk})s_{k}$
are responsible for scrambling \cite{vhecke2021, teunisse}.}

{To realize a Preisach scaffold defined by hysteron switching thresholds $h_i^\pm$ with interacting spins, 
consider the spin model with parameters:
\begin{eqnarray}
u_i^+=u_i^-=u_i&=& h_i^+~,\\
c_{ij}&=&-\sigma_j=h_i^- - h_i^+~.
\end{eqnarray}
We note that the 'column-wise' form of the second equation makes sure that the scrambling term is zero.
Substituting these parameters in the scaffold design inequalities
(Eq.~\ref{des1}-\ref{des2})  yields:
\begin{eqnarray}
U_i^+(S) - U_j^+(S) = h_i^+ - h_j^+  &>& 0~,\\
U_i^-(S) - U_j^-(S) =h_i^+ -\sigma_i - h_j^+ -\sigma_j = h_i^--h_j^- &>& 0~,
\end{eqnarray}
which are the inequalities for a Preisach system. Hence, interacting spins can form all Preisach scaffolds.}

{We now show that, through an extension of the mapping shown above, interacting spins can realize any scaffold realizable by pairwise interacting hysterons. We map any given
hysteron parameters $h_i^\pm$ and $\tilde{c}_{ij}$ to spin parameters:
\begin{eqnarray}
u_i^+=u_i^-=u_i&=& h_i^+~, \label{triv}\\ 
c_{ij}&=&-\sigma_j+\tilde{c}_{ij}~. \label{shi}
\end{eqnarray}
Substituting these parameters in the scaffold design inequalities
(Eq.~\ref{des1}-\ref{des2}) {-- where we note}  
that $h_i^+ -\sigma_i=h_i^-$, and that the sum {$(\tilde{c}_{ik} - \tilde{c}_{jk})s_{k}$ is invariant under the 'column-wise' translations stipulated by Eq.~\ref{shi} --} we find that for the spin model
\begin{eqnarray}
U_i^+(S) - U_j^+(S) &=& h_i^+ - h_j^+  - \sum_{k \neq i, j}(\tilde{c}_{ik} - \tilde{c}_{jk})s_{k} > 0~,\\
U_i^-(S) - U_j^-(S) &=& h_i^- - h_j^+ - \sum_{k \neq i, j}(\tilde{c}_{ik} - \tilde{c}_{jk})s_{k} > 0~ ,
\end{eqnarray}
which are manifestly the design inequalities for the hysteron model. Hence, the scaffolds of any hysteron model can be mapped to the scaffold of a spin model using the spin parameters 
Eq.~(\ref{triv}-\ref{shi}). }

\begin{figure}[t]
\centering
\includegraphics[width=.6\textwidth]{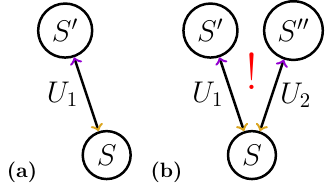}
\caption{{Illustration of the impossibility of nontrivial scaffolds in spin systems without avalanches}. a) A single reversible spin flip for a t-graph without avalanches, occurring at $U_1$. b) Posited part of a t-graph {for a spin system} without avalanches, {consisting of} two transitions and three states. {Since there are two up transitions from state $S'$ this t-graph is inconsistent, as indicated by the exclamation mark; thus, as stated in the main text, it is not possible to realize nontrivial t-graphs in spin systems without avalanches.}}
\label{fig:M1}
\end{figure}

\paragraph{Nontrivial Scaffolds produce avalanches.---}

Just from looking at the scaffold, we cannot distinguish  interacting hysteron systems from interacting spin systems. There is a crucial difference, however, between the t-graphs of interacting spins and interacting hysterons: for spin systems, nontrivial scaffolds {\em lead to avalanches in the t-graphs}. 

{We show this via proof by contradiction: if a t-graph for an interacting spin system is avalanche-free, it must be trivial.} Consider 
an up transition $S\uparrow S'$ that occurs at switching thresholds $U_1$; as the t-graph is avalanche-free, there is also a down transition
$S'\downarrow S$ at switching thresholds $U_1$ \mht{(Fig.~\ref{fig:M1}a)}. 
Now suppose that there is another state
$S''$, so that $S''\downarrow S$ at switching thresholds $U_2$; this implies that the up transition  $S\uparrow S'' $ occurs at switching threshold $U_2$. If $U_2>U_1$, this implies an avalanche
$S''\downarrow S \uparrow S'$; and if $U_2<U_1$, then the transition $S\uparrow S'$ cannot occur \mht{(Fig.~\ref{fig:M1}b)}. In other words, without avalanches, t-graph topologies of interacting spins are free of {'branches'}. As a result, for spins, t-graph topologies without avalanches are of the form $(\dots 000) \leftrightarrow (\dots 001)\leftrightarrow (\dots 011) \leftrightarrow \dots$, up to relabeling of the spins. \mht{By extension}, t-graphs {that have} any other scaffold {must} feature avalanches.

\subsection{Spins pairs can effectively form a 
single hysteron.}
\label{ss:spinpair}

The up and down transitions between two spin states that differ by the phase of one spin are equal, but for states that are connected by avalanches, this no longer needs to be true. 
As a consequence, the behavior of a single hysteron can be 
mimicked by pairs of interacting spins that 
form t-graphs with only two states, $(00)$ and $(11)$, connected by avalanches. There are (up to relabeling symmetry) two scaffolds that allow to create such t-graphs (Fig.~\ref{fig:m2}a-b). 
In the first, the intermediate state is the same for both avalanches, and in the second, the intermediate state is different.
Below we show that the first case can be captured by spin pairs with symmetric interactions, whereas the second case requires (strongly) asymmetric interactions\footnote{A brief version of this work was included as supplementary information in~\cite{colintoggleron2025}.}. 
While these two constructions are not exhaustive - we choose a particular parametrization for the asymmetric case, and constructions involving more than two spins are also possible - they highlight the simplest manner in which individual hysterons can be mimicked by interacting spins. We note that the asymmetric construction is a bit more involved; yet, 
when mimicking interacting hysterons by interacting pairs of spins, the symmetric construction leads to issues with race conditions, {while these} can be {avoided for}
asymmetrically interacting spin pairs.

\begin{figure}\label{fig:spin_constructions}[t]
\centering
\includegraphics[width=.8\textwidth]{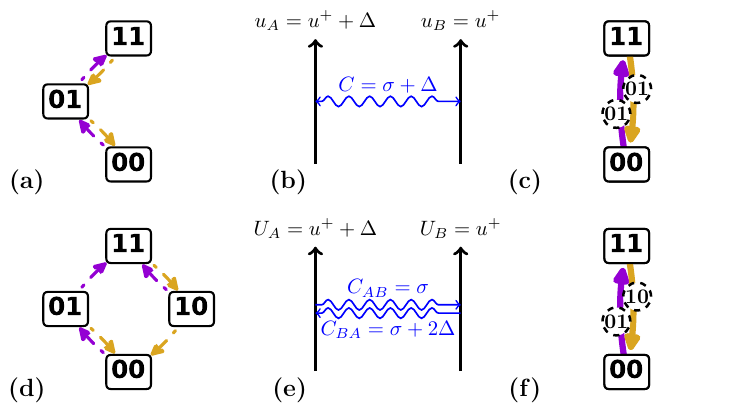}
\caption{Two constructions to replicate hysteron behavior using  interacting spins. (a) {Trivial}  scaffold for two spins. (b) Schematic of a symmetrically coupled pair of spins that realizes the trivial scaffold.
(c) T-graph for symmetrically coupled spins featuring
hysteretic avalanche transitions between $(00)$ and $(11)$ with the same intermediate state {$(01)$}. {(d) Nontrivial scaffold for two spins.}
(e) Schematic of an asymmetrically coupled pair of spins that realizes the {nontrivial} scaffold.
(f) T-graph for asymmetrically coupled spins featuring 
hysteretic avalanche transitions between $(00)$ and $(11)$ with different intermediate states {$(01)$ and $(10)$}.
}\label{fig:m2}
\end{figure}

\paragraph{Symmetric Spin Pair.---}
We consider two spins, $A$ and $B$, and aim to find switching thresholds and interactions such that the t-graph consists of two avalanches that connect $(00)$ and $(11)$, so that we can associate
$(00)$ and $(11)$ with hysteron states {$0$} and {$1$}
with switching thresholds $u^+,u^-$.
We denote the hysteron parameters with lowercase symbols
$u_i^\pm$ and $c_{ij}$, and the spin parameters with capitals
$U_{i(A/B)}$ and $C_{i(A/B),j(A/B)}$.
We take $U_A - U_B := \Delta > 0$ and consider positive symmetric interactions $C_{AB} = C_{BA} = C>0$. 
The scaffold then involves three states, with {the transitions}
$(00) \leftrightarrow (01)$ occurring at $U_B$, and 
$(01) \leftrightarrow (11)$  occurring at $U_A-C=U_B+\Delta-C$ 
(Fig.~\ref{fig:m2}a). For $C>\Delta$, the t-graph features two avalanches, with the up transition $(00)\rightarrow(11)$ {occurring} at
$U_B$, and the down transition $(11)\rightarrow(00)$ at $U_B+\Delta-C$. Note that in this construction, the up transition is initiated by spin $B$ switching up, yet the down transition is initiated by spin $A$ switching down.

Taking $0<\Delta<C$, we can identify the 
spin states $(00)$ and $(11)$ with {the states $0$ and $1$ of a hysteron, which has switching thresholds $u^\pm$}(Fig.~\ref{fig:m2}c): 
\begin{eqnarray}
u^+ &=& U_B~,  \label{eq24}\\
u^-&=& U_B+\Delta-C~.\label{eq25}
\end{eqnarray}
We stress that at the level of a single hysteron, 
the parameter $\Delta$ can be chosen arbitrarily, but this parameter will play a role when modeling interacting hysterons
(section \ref{ss:rep}).

\paragraph{Asymmetric Spin Pair.---}
Hysterons can also be modeled using a scaffold with four states (Fig.~\ref{fig:m2}b).
As above, we take $U_A - U_B := \Delta > 0$, where $\Delta$ can be freely chosen. Realizing this scaffold
requires sufficiently strong asymmetric interactions such that 
$C_{AB}> C_{BA}+\Delta$; for convenience, we take $C_{AB}$
and $C_{BA}$ {to be} positive.
The switching thresholds for the four scaffold transitions are:
$(00)\leftrightarrow(01)$ at $U_B$;
$(01)\leftrightarrow(11)$ at $U_A-C_{AB} = U_B+\Delta-C_{AB}$;
$(11)\leftrightarrow(10)$ at $U_B-C_{BA}$;
$(10)\leftrightarrow(00)$ at $U_A=U_B+\Delta$.
Then, to have a $(00)\rightarrow (11)$ avalanche, we require $C_{AB}>\Delta$; the {transition} $(11)\rightarrow (00)$ is guaranteed for positive $C_{BA}$ (an example of the intertwining of scaffolds and avalanches).
Note that, in contrast to the symmetrically coupled construction, here the up and down avalanches are both initiated by spin $B$ switching up and down respectively.

We can again identify the 
spin states $(00)$ and $(11)$ with hysteron states {$0$} and {$1$}, that have switching thresholds {$u^\pm$}  (Fig.~\ref{fig:m2}d): 
\begin{eqnarray}
u^+ &=& U_B~, \label{up2}\\
u^-&=& U_B-C_{BA}~. \label{um2}
\end{eqnarray}
To pick the spin parameters, we first fix $U_B$ and $C_{BA}$ following Eq.~\ref{up2}-\ref{um2},
{then} take arbitrary $\Delta>0$, and {finally} choose a large value of $C_{AB}$
such that $C_{AB}> C_{BA}+\Delta$ to satisfy the constraints for the scaffold; for definiteness, we set $C_{AB} = C_{BA}+2\Delta$.

\begin{figure}
\centering
\includegraphics[width=.9\textwidth]{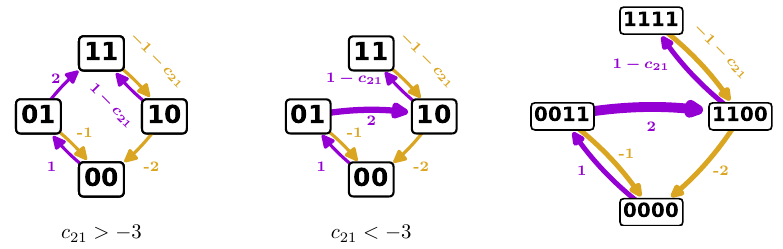}
\caption{Replicating a system of interacting hysterons using spin pairs. a-b) T-graphs for two-hysteron systems
as discussed in the main text, for $c_{21}>-3$ (a) and 
$c_{21}<-3$ (b)
c) T-graph for $n=4$ spins which mimic the $n=2$ t-graph in panel (b).}
\label{fig:m3}
\end{figure}

\subsection{Reproducing hysteron t-graphs using interacting spin pairs}\label{ss:rep}
We now discuss how the behavior of
$n$ interacting hysterons can be captured by $2n$ interacting spins.  {We first translate each hysteron to a spin pair}, e.g., $1A,1B$ models hysteron 1, so that a hysteron state $S=(s_1,s_2,s_3)$ maps to a spin state $\tilde{S}=(s_1,s_1,s_2,s_2,s_3,s_3)$. We {refer to spin states of this form}, where each spin pair is in state $(00)$ or $(11)$, {as 'pure'}. {We} use interactions
within each pair, as outlined above, to model the bare hysterons. This means that a single hysteron flip maps to an avalanche of two spin flips within a spin pair, with an intermediate state where one spin pair is in state $(01)$ or $(10)$
(Fig.~\ref{fig:m2}e-f). 

{Having translated the bare hysterons to spin pairs}, 
we {then} introduce interactions between these pairs of spins to model the hysteron
interactions. Such a mapping is successful when {\em (i)} any
hysteron transition in the t-graph
$S\rightarrow S'$ - including avalanches -
maps to spin avalanche transitions between pure states $\tilde{S}\rightarrow \tilde{S}'$; {\em(ii)} mixed spin states (where one or more spin pairs are not in state $(00)$ or $(11)$) {are never stable;} and {\em(iii)} no race conditions occur in the spin system{, except when these are inherited from race conditions in the original} hysteron system.

As discussed above, there are at least two strategies to model the bare hysterons{; as we show below,} there are also multiple strategies to model the hysteron interactions. {We show that}
for collections of $n$ hysterons whose t-graphs do not exhibit avalanches, several parametrizations of $2n$ spins capture those t-graphs{. Yet,} when the hysteron t-graph feature avalanches, one has to be careful to avoid race conditions in the corresponding spin systems. {These race conditions can be 'internal', where both spins in pair $i$ lose stability at the same time, or 'external', where} the flip of one spin in pair $i$ already destabilizes spins in pair $j$. {We show a specific mapping which avoids both types of race conditions, thus allowing $n$ interacting hysterons to be faithfully captured by $2n$ interacting spins}.

We illustrate the emergence of race conditions in pairs of spins that represent interacting hysterons with a concrete example, before discussing the general case.
Suppose that we have a pair of hysterons with bare switching thresholds:
\begin{equation} \label{eq:ui}
(u_1^+, u_2^+, u_1^-, u_2^-) = (2, 1, -2, -1)
\end{equation}
and interactions 
\begin{equation} \label{eq:cij}
c_{ij} = 
\begin{pmatrix}
0 & 0\\
c_{21} & 0\\
\end{pmatrix}~,
\end{equation}
where we assume $c_{21}<0$. Depending on the value of 
{$c_{21}$}, we find two hysteron t-graphs ---
for {$c_{21}>-3$}, the t-graph has no avalanches, whereas
for {$c_{21}<-3$}, the t-graph features the {avalanche}
$(01)\xrightarrow{(11)}(10)$ (Fig.~\ref{fig:m3}a-b).
Note that this system is free of race conditions, {since race conditions can only occur in hysteron systems of three or more elements} \cite{baconnier2025dynamic}.
We will now consider whether race conditions emerge when we map this {two-hysteron} system to two interacting spin pairs.

\paragraph{Race conditions for mapping I.---}
First, we consider mapping I, where each hysteron is represented by a symmetrically coupled spin pair, and hysteron interactions are evenly distributed over the $A$ and $B$ spins. Using Eq.~(\ref{eq24}-\ref{eq25}), and {taking} $\Delta$ equal for each pair, spin pair 1 has parameters
\begin{eqnarray} \label{mapI1}
U_{1A}&=&2+\Delta~,\\
U_{1B}&=&2~,\\
C_{A1,B1}=C_{B1,A1}&=&4+\Delta~,\\
\end{eqnarray}
and spin pair 2 has parameters:
\begin{eqnarray}  \label{mapI2}
U_{2A}&=&1+\Delta~,\\
U_{2B}&=&1~,\\
C_{A2,B2}=C_{B2,A2}&=&2+\Delta~.
\end{eqnarray}

We then map the hysteron interaction $c_{21}$ to interactions
between spin pairs 1 and 2. We take 
\begin{equation}
C_{2A,1A} = C_{2A,1B} = C_{2B,1A} = C_{2B,1B} = c_{21}/2~,
\end{equation}
so that spins $1A$ and $1B$ contribute equally to the 
shift of the switching thresholds of 
spins $2A$ and $2B$. 
The coupling matrix for the four spin system ({where} we order rows and colums as $1A,1B,2A,2B$) thus becomes
:
\begin{equation}  \label{mapI3}
C=\begin{pmatrix}
0&4 +\Delta&0&0\\
4+\Delta&0&0&0\\
c_{21}/2&c_{21}/2&0&2+\Delta\\
c_{21}/2&c_{21}/2&2+\Delta&0
\end{pmatrix}~.
\end{equation}
We note that this matrix has a clear $2\times 2$ block structure, with the center blocks containing internal interactions, and the off-diagonal blocks containing the {external} interactions -- {i.e., interactions between spin pairs, which are inherited from the original hysteron system}. {We can generically} write the spin interaction matrix as
\begin{equation}  
C=\begin{pmatrix}
0&\sigma_1 +\Delta&c_{12}/2&c_{12}/2\\
\sigma_1+\Delta&0&c_{12}/2&c_{12}/2\\
c_{21}/2&c_{21}/2&0&\sigma_2+\Delta\\
c_{21}/2&c_{21}/2&\sigma_2+\Delta&0
\end{pmatrix}~,
\end{equation}
where $\sigma_i
$ denotes the hysteron span $u_i^+-u_i^-$.

We now vary $c_{21}$ to illustrate how race conditions can emerge {for} mapping I. {We consider} how the 
hysteron transitions $(01)\rightarrow (11)$,  occurring 
at $u_1^+-c_{12}=2$, and
$(11)\rightarrow (10)$, 
occurring 
at $u_2^- - c_{21}=-1-c_{21}$,
map to spin avalanche transitions 
$(0011)\xrightarrow{(0111)} (1111)$ and 
$(1111)\xrightarrow{(1101)} (1100)$ -- here
$(0111)$ and $(1101)$ denote intermediate states. To {show how race conditions emerge}, we calculate the 
switching thresholds for these states (Table.~\ref{tab:gen}) and work out {their} explicit values as function of $c_{21}$ (see Appendix).

\mvh{\begin{table}[t]
\centering
\begin{tabular}{|c|c|c|c|c|}
\hline
$S$ & $U_{1A}(S)$  & $U_{1B}(S)$ & $U_{2A}(S)$ & $U_{2B}(S)$ \\ 
 &2+$\Delta$ & 2 &1+$\Delta$ & 1 \\
   & $-(4+\Delta)S_{1B}$ 
   & $-(4+\Delta)S_{1A} $ 
   & $-\frac{c_{21}}{2}S_{1A} -\frac{c_{21}}{2}S_{1B}$ 
   & $-\frac{c_{21}}{2}S_{1A} -\frac{c_{21}}{2}S_{1B}$ \\
& & &$-(2+\Delta)S_{2B}$ &$-(2+\Delta)S_{2A}$  \\
\hline
(0011)     & $2+\Delta$    & 2     &-1      &$-1-\Delta$    \rule{0pt}{10pt} \\ 
(0111)     &   -2   &  2    & $-1-\frac{c_{21}}{2}$     & $-1-\Delta-\frac{c_{21}}{2} $     \rule{0pt}{10pt}\\ 
(1111)     &   -2   &  $-2-\Delta$    & $-1-c_{21}$     & $-1-\Delta -c_{21} $     \rule{0pt}{10pt}\\
(1101) &  -2  & $-2-\Delta$     & $-1-c_{21}$     & $1-c_{21}$           \rule{0pt}{10pt} \\ 
(1100)  & -2  & $-2-\Delta$     & $1+\Delta-c_{21}$     & $1-c_{21}$          \rule{0pt}{10pt}  \\ \hline 
\end{tabular}
\caption{Switching thresholds for selected states for spin parametrization I} \label{tab:gen}
\end{table}}

We first consider {how} mapping I {applies to} the avalanche-free
hysteron t-graph obtained for 
$c_{21}=-1$ (Fig.~\ref{fig:m3}a).
The specific values of the switching thresholds are tabulated in Table.~\ref{tab:c-1}. 
Starting out from $(0011)$, we see that this state becomes unstable when $U$ is increased above 2, which we denote as
$U=2^+$. This {instability} triggers the spin flip 
$(0011)\rightarrow(0111)$; in $(0111)$ spin $S_{1A}$ is unstable, {so the avalanche reaches} state $(1111)$, which is stable at $U=2^+$. Hence, the hysteron transition
$(01)\rightarrow(11)$ is captured by the spin avalanche
$(0011)\xrightarrow{(0111)} (1111)$.
Similarly, starting from $(1111)$ and lowering $U$, 
spin 2A flips down when $U$ falls below 0, leading to state $(1101)$,
where spin 2B is unstable and flips up, leading to spin state $(1100)$ which is stable at $U=0^-$. 
Hence, the hysteron transition
$(11)\rightarrow(10)$ is captured by the spin avalanche
$(1111)\xrightarrow{(1101)} (1100)$.
We note that these transitions are independent of the value of $\Delta$, as long as {$\Delta > 0$}.
Similarly, all other hysteron transitions of this system are captured by
internal spin avalanches, i.e. avalanches that only involves spins within a single pair (Fig.~\ref{fig:m3}c).  This example illustrates {the} general point {that}
mapping I allows {a conversion from}
avalanche-free hysteron t-graphs to interacting spin pairs{, which} only feature 'internal' avalanches ({i.e.}, within each spin pair).

However, when the hysteron t-graph features avalanches, race conditions can arise in the corresponding spin system. To see this, we now {choose} $c_{21}=-4$, {which leads to} a hysteron
t-graph with avalanche $(01)\xrightarrow{(11)}(10)$ (Fig.~\ref{fig:m3}b). {For this system, mapping I} yields the switching thresholds in Table.~\ref{tab:c-4}.
As before, starting out from $(0011)$, we see that this state becomes unstable when $U$ is increased above 2, which  triggers the spin flip 
$(0011)\rightarrow(0111)$, and that 
in $(0111)$ spin $S_{1A}$ is unstable leading to state $(1111)$. However, in state $(1111)$ at $U=2^+$, 
both spin 2A and 2B are unstable for small $\Delta$. If we choose $\Delta>1$, then only spin 2A is unstable and flips down, leading to state (1101), where now only spins 2B is unstable, leading to state (1100) that is stable at $U=2+\varepsilon$. This illustrates the general point that using mapping I, avalanches in the hysteron system can cause race conditions in the spin system. For this specific case, {the race condition} can be resolved by taking $\Delta>1$ --- intuitively, by increasing $\Delta$ we increase the magnitude of the interactions between spins in a pair, thus making {the link between the spins 'stronger'}.
{Race conditions which are 'internal', meaning that both spins in the same pair become unstable at the same time, can thus be resolved by increasing $\Delta$.} 

{However}, mapping I leads to irresolvable {'external'} race conditions when $c_{ij}$ is decreased further. Consider $c_{21}=-8$, {which leads} to  switching thresholds as in Table.~\ref{tab:c-8}. 
Again, state $(0011)$ becomes unstable when $U$ is increased above 2 leading to  
$(0011)\rightarrow(0111)$. However, in this state spin 1A is unstable to flipping up, and spin 2A is unstable to flipping down. Crucially, as both thresholds are independent of $\Delta$, 
{an increase in $\Delta$ cannot help to} resolve this race condition.

To understand the reason for the emergence of this race condition in the spin model, {we note} that in mapping I, the {flips} of spins A and B {in} hysteron $j$ both impact the switching thresholds of all
spins A and B in pairs $i$ where $c_{ij}\ne 0$. Hence, not only the switching  thresholds of the spins in the pure states are affected, but also those in the 'mixed' intermediate states. Therefore, when hysteron $j$ flips by  
flipping spins $jA$ and $JB$ in an avalanche,
the first spin flip can already destabilize multiple spins
in a mixed state,
leading to unwanted race conditions.
{For hysteron t-graphs which have avalanches, mapping I can thus produce} spin systems where race conditions
cannot be resolved by increasing $\Delta$.

\paragraph{Mapping II.---}
We construct a mapping where the flipping of the first spin of pair $j$
does not couple to any spins except the other spin in pair $j$,
thus avoiding race conditions during the intermediate stage of flipping any spin pair. We can realize this by making two changes to mapping I. First, we note that in the symmetric spin pair, spin B is the first one to flip up, where spin A is the first one to flip down - so there is no consistent switching order. Instead we therefore use the asymmetric spin pair, where spin B is {always} the first one to flip up or down (Fig.~\ref{fig:m2}f). 
We then make sure that 
the B spins only couple to A spins in the same spin pair, leading to an 'internal' avalanche, with all the hysteron interactions $c_{ij}$ encoded in interactions from spin $jA$ to both 
$iA$ and $iB$, so that no race conditions can occur in the mixed/intermediate spin states. {We call this mapping II.}

We illustrate mapping II for the specific hysteron system given by Eqs.~\ref{eq:ui}-\ref{eq:cij}. For spin pair 1 we take 
\begin{eqnarray} \label{mapI33}
U_{1A}&=&2+\Delta~,\\
U_{1B}&=&2~,\\
C_{A1B1}&=&4 + 2 \Delta~,\\
C_{B1A1}&=&4~,\\
\end{eqnarray}
and for spin pair 2:
\begin{eqnarray}  \label{mapI4}
U_{2A}&=&1+\Delta~,\\
U_{2B}&=&1~,\\
C_{A2B2}&=&2 + 2 \Delta~,\\
C_{B2A2}&=&2~,\\
\end{eqnarray}
To capture the hysteron interactions $c_{21}$ we take
\begin{equation}
C_{A2,A1}=C_{B2,A1}=c_{21}~,
\end{equation}
and 
to make sure that 
the B spins {have no external coupling, instead only being coupled to} A spins in the same spin pair, we set:
\begin{equation}
C_{A2,B1}=C_{B2,B1}=0~.\label{eq:zero}
\end{equation}
Together, Eq.~\ref{mapI33}-\ref{eq:zero} define mapping II. Its interaction matrix is:
\begin{equation}  \label{mapCII}
C=\begin{pmatrix}
0&4+2 \Delta &0&0\\
4&0&0&0\\
c_{21}&0&0&2+2\Delta\\
c_{21}&0&2&0
\end{pmatrix}~.
\end{equation}

We now consider the pathways for the same t-graph as before, for $c_{21}=-1,-4$ and $-8$. The switching thresholds for this specific system are
\begin{table}[h]
\centering
\begin{tabular}{|c|c|c|c|c|}
\hline
$S$ & $U_{1A}(S)$  & $U_{1B}(S)$ & $U_{2A}(S)$ & $U_{2B}(S)$ \\ 
 &2+$\Delta$ 
  & 2 &1+$\Delta$ & 1 \\
   & $-(4+2\Delta)S_{1B}$ 
   & $-4S_{1A} $ 
   & $-c_{21}S_{1A} $ 
   & $-c_{21}S_{1A}$  \\
& & &$-(2+2\Delta)S_{2B}$ &$-2S_{2A}$  \\
\hline
(0011)     & $2+\Delta$    & 2     &$-1-\Delta$      &$-1$    \rule{0pt}{10pt} \\ 
(0111)     &   $-2-\Delta$   &  2    & $-1-\Delta$     & $-1 $     \rule{0pt}{10pt}\\ 
(1111)     &   $-2-\Delta$   &  $-2$    & $-1-\Delta-c_{21}$     & $-1-c_{21}$      \rule{0pt}{10pt}\\
(1110) &  $-2-\Delta$  & $-2$     & $1+\Delta-c_{21}$     & $-1-c_{21}$           \rule{0pt}{10pt} \\ 
(1100)  & $-2-\Delta$  & $-2$     & $1+\Delta-c_{21}$     & $1-c_{21}$          \rule{0pt}{10pt}  \\ \hline 
\end{tabular}
\caption{Switching thresholds for selected states for spin parametrization II} \label{tab:genII}
\end{table}

We first consider the spin transitions for $c_{21}=-1$, 
where the t-graph of the hysteron system has no avalanches. Calculating the switching thresholds (Table.~\ref{tab:genII-1}), we find that for any $\Delta>0$ the spin system has an internal avalanche $(0011)\xrightarrow{(0111)}(1111)$ at $U=2+$, and
an internal avalanche $(1111)\xrightarrow{(1110)}(1100)$ at $U=0-$, mimicking the 
$(01)\rightarrow(11)$ and 
$(11)\rightarrow(10)$ transitions of the hysteron model; no} race conditions occur.
Second, for $c_{21}=-4$ the hysteron model features the avalanche 
$(01)\xrightarrow{(11)}(10)$ at $U=2+$. The spin switching thresholds (Table.~\ref{tab:genII-4}) lead to the 
spin avalanche
$(0011)\xrightarrow{(0111)(1111)(1110)}(1100)$. Race conditions in state $(1111)$ can be avoided by taking $\Delta>1$.
Finally, for $c_{21}=-8$, the spin switching thresholds (Table.~\ref{tab:genII-8}) lead to the 
spin avalanche
$(0011)\xrightarrow{(0111)(1111)(1110)}(1100)$ where race conditions in state $(1111)$ can be avoided by taking $\Delta>5$. Hence, using mapping II, for large enough $\Delta$ race conditions can be avoided and even hysteron systems with avalanches can be faithfully mapped to interacting spin pairs.

We now argue why mapping II, in general, provides a faithful mapping from a hysteron system to a spin system.
First, for any pure state the switching thresholds of spins $s_{iB}$ correspond to the switching thresholds of hysteron $s_i$
and the 
switching thresholds of spins $s_{iA}$ are larger or smaller than those of spins $s_{iB}$ if 
$s_{iB}=0$ or $s_{iB}=1$. As a consequence, the range of stability of any hysteron state is identical to the 
range of stability of the corresponding spin state, 
and when a state is destabilized and hysteron $i$ flips, 
the $iB$ spin will always switch first.
Second, once a $B$ spin flips, only the corresponding $A$ spin threshold changes and is forced to flip, as long as $\Delta$ is large enough. Hence, no race conditions occur 
for large enough $\Delta$. As a result, mapping II (which for general $c_{ij}$ can easily be written down following the block structure of $C$) correctly maps any well-behaved interacting hysteron system to a well behaved binary spin system.

\section{Outlook}

We have studied the hysteron model in for large and small
hysteron spans, and established a link between interacting hysterons and interacting spins. {We have shown that for hysteron systems where the mean span is large, the only allowed avalanches are monotonic avalanches, where all flips are either up or down. For large-span systems with purely negative interactions $c_{ij}$, avalanches are entirely suppressed. By contrast, in the 'spin' limit - where the hysteron span approaches zero - avalanches are required in order to realize nontrivial t-graphs. By mapping hysterons to pairs of spins with internal avalanches, we can mimic a system of $n$ hysterons with $2n$ spins, and we show one specific mapping that avoids the occurence of spurious race conditions.}

We {close this discussion by making two comments: first, on the gauge symmetry of the hysteron model paramaters, and second, on the relation between hysterons and spins. To start with, because} the t-graphs of hysteron models only depend on the order of the switching thresholds, the pairwise interacting model features  a gauge symmetry of the form
$(u_i^\pm,c_{ij}) \leftrightarrow (\lambda~u_i^\pm,\lambda~c_{ij})$,
where $\lambda>0$ \cite{vhecke2021,teunisse}. Roughly speaking, the scatter {$u_i^+ - u_j^+$ and $u_i^- - u_j^-$ in the bare thresholds}, the {mean} span, and the strength of the interactions are important metaparameters that control the typical t-graphs that can be found. Hence, the large-span limit is equivalent to the limit where the spans are fixed and {the interactions $c_{ij}$, as well as the scatter in the up and down thresholds, approach zero. In other words, this is a  system of near-identical hysterons with weak coupling \cite{aref}. Similarly, the 'spin' limit is equivalent to a system with large coupling $c_{ij}$ and large scatter in the bare up and down switching thresholds. We note, however, that in this case we have the restriction $u_i^+ - u_j^+ \approx u_i^- - u_j^-$ for the scatter; in other words, not all systems which have large coupling and scatter behave like a spin system.}

We finally comment on our mapping from the hysterons to spins. We {stress that, while a system of interacting systems can be faithfully mapped to twice as many interacting spins, it remains meaningful to view hysterons as being distinct from spin systems}. {Namely,} we note that this mapping requires spins to be paired in a peculiar{, asymmetric} manner, and that their internal interactions $\Delta$ need to dominate all
other interactions. {Thus, while this mapping shows that hysterons can be understood via a spin construction, it also makes explicit how hysteron behaviour is markedly different from} generically coupled spins. Nevertheless, it may be interesting to consider whether, in some physical systems, hysterons {could naturally emerge} from pairs of strongly interacting nearby spins. Similarly, it would be interesting to consider {'hysteron-like' spin systems} in which $\Delta$ is not large enough, so that the hysterons can be 'broken apart' by strong interactions.
Finally, {as} hysterons can be seen as {being} composed of spins, this suggests {that we might} consider a hierarchy of composite elements. This hierarchy may also feature more complex elements, such as binary elements that return to their initial state after two driving 
cycles \cite{toggleron}.

\ack{We thank Colin Meulblok and Sourav Roy for fruitful discussions {regarding spin systems and large-span systems, respectively}.}

\funding{MT and MvH acknowledge funding from European Research Council Grant ERC-101019474.}

\appendix
\section{Switching thresholds}

Below we provide the explicit values of the spin thresholds corresponding mapping I and II as discussed in the main text. Specifically, we consider a hysteron system
where $(u_1^+, u_2^+, u_1^-, u_2^-) = (2, 1, -2, -1)$ and
$c_{12}=0$. We vary $c_{21}$ and consider the hysteron transitions $(01)\rightarrow(11)$ and
$(11)\rightarrow(10)$, which for strongly negative
$c_{21}$ form the avalanche $(01)\xrightarrow{(11)}(10)$.

\mvh{\begin{table}[h]
\centering
\begin{tabular}{|c|c|c|c|c|}\hline
$S$ & $U_{1A}(S)$  & $U_{1B}(S)$ & $U_{2A}(S)$ & $U_{2B}(S)$ \\ 
 &2+$\Delta$ & 2 &1+$\Delta$ & 1 \\
   & $-(4+\Delta)S_{1B}$ 
   & $-(4+\Delta)S_{1A} $ 
   & $-\frac{c_{21}}{2}S_{1A} -\frac{c_{21}}{2}S_{1B}$ 
   & $-\frac{c_{21}}{2}S_{1A} -\frac{c_{21}}{2}S_{1B}$ \\
& & &$-(2+\Delta)S_{2B}$ &$-(2+\Delta)S_{2A}$  \\
\hline
(0011)     & $2+\Delta$    & 2     &-1      &$-1-\Delta$    \rule{0pt}{10pt} \\ 
(0111)     &   -2   &  2    & $-1/2$     & $-1/2-\Delta   $ \rule{0pt}{10pt}\\ 
(1111)     &   -2   &  $-2-\Delta$    & 0     & $-\Delta  $     \rule{0pt}{10pt}\\
(1101) &  -2  & $-2-\Delta$     & $0$     & $2$           \rule{0pt}{10pt} \\ 
(1100)  & -2  & $-2-\Delta$     & $2+\Delta $    & 2          \rule{0pt}{10pt}  \\ \hline 
\end{tabular} 
\caption{Switching thresholds for selected states for spin parametrization I and $c_{21}=-1$.}\label{tab:c-1}
\end{table}}

\mvh{\begin{table}[h]
\centering
\begin{tabular}{|c|c|c|c|c|}\hline
$S$ & $U_{1A}(S)$  & $U_{1B}(S)$ & $U_{2A}(S)$ & $U_{2B}(S)$ \\ 
 &2+$\Delta$ & 2 &1+$\Delta$ & 1 \\
   & $-(4+\Delta)S_{1B}$ 
   & $-(4+\Delta)S_{1A} $ 
   & $-\frac{c_{21}}{2}S_{1A} -\frac{c_{21}}{2}S_{1B}$ 
   & $-\frac{c_{21}}{2}S_{1A} -\frac{c_{21}}{2}S_{1B}$ \\
& & &$-(2+\Delta)S_{2B}$ &$-(2+\Delta)S_{2A}$  \\
\hline
(0011)     & $2+\Delta$    & 2     &-1      &$-1-\Delta$    \rule{0pt}{10pt} \\ 
(0111)     &   -2   &  2    & $1$     & $1-\Delta   $ \rule{0pt}{10pt}\\ 
(1111)     &   -2   &  $-2-\Delta$    & 3     & $3-\Delta  $     \rule{0pt}{10pt}\\
(1101) &  -2  & $-2-\Delta$     & $3$     & $5$           \rule{0pt}{10pt} \\ 
(1100)  & -2  & $-2-\Delta$     & $5+\Delta $    & 5          \rule{0pt}{10pt}  \\ \hline 
\end{tabular} 
\caption{Switching thresholds for selected states for spin parametrization I and $c_{21}=-4$.}\label{tab:c-4}
\end{table}}

\mvh{\begin{table}[h]
\centering
\begin{tabular}{|c|c|c|c|c|}\hline
$S$ & $U_{1A}(S)$  & $U_{1B}(S)$ & $U_{2A}(S)$ & $U_{2B}(S)$ \\ 
 &2+$\Delta$ & 2 &1+$\Delta$ & 1 \\
   & $-(4+\Delta)S_{1B}$ 
   & $-(4+\Delta)S_{1A} $ 
   & $-\frac{c_{21}}{2}S_{1A} -\frac{c_{21}}{2}S_{1B}$ 
   & $-\frac{c_{21}}{2}S_{1A} -\frac{c_{21}}{2}S_{1B}$ \\
& & &$-(2+\Delta)S_{2B}$ &$-(2+\Delta)S_{2A}$  \\
\hline
(0011)     & $2+\Delta$    & 2     &-1      &$-1-\Delta$    \rule{0pt}{10pt} \\ 
(0111)     &   -2   &  2    & $3$     & $3-\Delta   $ \rule{0pt}{10pt}\\ 
(1111)     &   -2   &  $-2-\Delta$    & 7     & $7-\Delta  $     \rule{0pt}{10pt}\\
(1101) &  -2  & $-2-\Delta$     & $7$     & $9$           \rule{0pt}{10pt} \\ 
(1100)  & -2  & $-2-\Delta$     & $9+\Delta $    & 9          \rule{0pt}{10pt}  \\ \hline 
\end{tabular} 
\caption{Switching thresholds for selected states for spin parametrization I and $c_{21}=-8$.}\label{tab:c-8}
\end{table}}

\mvh{\begin{table}[h]
\centering
\begin{tabular}{|c|c|c|c|c|}
\hline
$S$ & $U_{1A}(S)$  & $U_{1B}(S)$ & $U_{2A}(S)$ & $U_{2B}(S)$ \\ 
 &2+$\Delta$ 
  & 2 &1+$\Delta$ & 1 \\
   & $-(4+2\Delta)S_{1B}$ 
   & $-4S_{1A} $ 
   & $-c_{21}S_{1A} $ 
   & $-c_{21}S_{1A}$  \\
& & &$-(2+2\Delta)S_{2B}$ &$-2S_{2A}$  \\
\hline
(0011)     & $2+\Delta$    & 2     &$-1-\Delta$      &$-1$    \rule{0pt}{10pt} \\ 
(0111)     &   $-2-\Delta$   &  2    & $-1-\Delta$     & $-1 $     \rule{0pt}{10pt}\\ 
(1111)     &   $-2-\Delta$   &  $-2$    & $-\Delta$     & $0$      \rule{0pt}{10pt}\\
(1110) &  $-2-\Delta$  & $-2$     & $2+\Delta$     & $0$           \rule{0pt}{10pt} \\ 
(1100)  & $-2-\Delta$  & $-2$     & $2+\Delta$     & $2$          \rule{0pt}{10pt}  \\ \hline 
\end{tabular}
\caption{Switching thresholds for selected states for spin parametrization II and $c_{21}=-1$.} \label{tab:genII-1}
\end{table}}

\mvh{\begin{table}[h]
\centering
\begin{tabular}{|c|c|c|c|c|}
\hline
$S$ & $U_{1A}(S)$  & $U_{1B}(S)$ & $U_{2A}(S)$ & $U_{2B}(S)$ \\ 
 &2+$\Delta$ 
  & 2 &1+$\Delta$ & 1 \\
   & $-(4+2\Delta)S_{1B}$ 
   & $-4S_{1A} $ 
   & $-c_{21}S_{1A} $ 
   & $-c_{21}S_{1A}$  \\
& & &$-(2+2\Delta)S_{2B}$ &$-2S_{2A}$  \\
\hline
(0011)     & $2+\Delta$    & 2     &$-1-\Delta$      &$-1$    \rule{0pt}{10pt} \\ 
(0111)     &   $-2-\Delta$   &  2    & $-1-\Delta$     & $-1 $     \rule{0pt}{10pt}\\ 
(1111)     &   $-2-\Delta$   &  $-2$    & $3-\Delta$     & $3$      \rule{0pt}{10pt}\\
(1110) &  $-2-\Delta$  & $-2$     & $5+\Delta$     & $3$           \rule{0pt}{10pt} \\ 
(1100)  & $-2-\Delta$  & $-2$     & $5+\Delta$     & $5$          \rule{0pt}{10pt}  \\ \hline 
\end{tabular}
\caption{Switching thresholds for selected states for spin parametrization II and $c_{21}=-4$.} \label{tab:genII-4}
\end{table}}

\mvh{\begin{table}[h]
\centering
\begin{tabular}{|c|c|c|c|c|}
\hline
$S$ & $U_{1A}(S)$  & $U_{1B}(S)$ & $U_{2A}(S)$ & $U_{2B}(S)$ \\ 
 &2+$\Delta$ 
  & 2 &1+$\Delta$ & 1 \\
   & $-(4+2\Delta)S_{1B}$ 
   & $-4S_{1A} $ 
   & $-c_{21}S_{1A} $ 
   & $-c_{21}S_{1A}$  \\
& & &$-(2+2\Delta)S_{2B}$ &$-2S_{2A}$  \\
\hline
(0011)     & $2+\Delta$    & 2     &$-1-\Delta$      &$-1$    \rule{0pt}{10pt} \\ 
(0111)     &   $-2-\Delta$   &  2    & $-1-\Delta$     & $-1 $     \rule{0pt}{10pt}\\ 
(1111)     &   $-2-\Delta$   &  $-2$    & $7-\Delta$     & $7$      \rule{0pt}{10pt}\\
(1110) &  $-2-\Delta$  & $-2$     & $9+\Delta$     & $7$           \rule{0pt}{10pt} \\ 
(1100)  & $-2-\Delta$  & $-2$     & $9+\Delta$     & $9$          \rule{0pt}{10pt}  \\ \hline 
\end{tabular}
\caption{Switching thresholds for selected states for spin parametrization II and $c_{21}=-8$.} \label{tab:genII-8}
\end{table}}

~\newpage
\printbibliography

\end{document}